\documentclass[twocolumn]{article}

\usepackage{geometry}
\usepackage{authblk}
\usepackage{cite}
\usepackage{lipsum}
\usepackage{amsmath}
\usepackage{amssymb}
\usepackage{graphicx}
\usepackage{gensymb}
\usepackage{tabularx}
\usepackage{float}
\usepackage{xcolor}
\usepackage{mathtools}
\usepackage{cancel}
\usepackage[version=4]{mhchem}
\usepackage[capitalise]{cleveref}
\usepackage[
singlelinecheck=false
]{caption}
\usepackage{threeparttable}
\usepackage{array}
\restylefloat{table}
\graphicspath{ {Images/} }

\newcommand{\SpK}[0]{\texttt{SpK}}
\title{Extension of the \SpK~atomic physics code to generate global equation of state data}

\author[*,1,2]{Adam R. Fraser}
\author[2]{A. J. Crilly}
\author[1,2]{N.-P. L. Niasse}
\author[1]{D. A. Chapman}
\author[1,2]{J. D. Pecover}
\author[2]{\newline S. J. O'Neill}
\author[2]{J. P. Chittenden}

\affil[1]{First Light Fusion Ltd., Unit 9/10 Oxford Industrial Park, Mead Road, Yarnton, OX5 1QU, Oxford, United Kingdom}
\affil[2]{Centre for Inertial Fusion Studies, The Blackett Laboratory, Imperial College, SW7 2AZ, London, United Kingdom}

\date{}
\onecolumn
\begin{document}
\twocolumn[
\maketitle

\begin{abstract}
Global microphysics models are required for the modelling of high-energy-density physics (HEDP) experiments, the improvement of which are critical to the path to inertial fusion energy. This work presents further developments to the atomic and microphysics code, \SpK, part of the numerical modelling suite of Imperial College London and First Light Fusion. We extend the capabilities of \SpK~to allow the calculation of the equation of state (EoS). The detailed configuration accounting calculations are interpolated into finite-temperature Thomas-Fermi calculations at high coupling to form the electronic component of the model. The Cowan model provides the ionic contribution, modified to approximate the physics of diatomic molecular dissociation. By utilising bonding corrections and performing a Maxwell construction, \SpK~captures the EoS from states ranging from the zero-pressure solid, through the liquid-vapour coexistence region and into plasma states. This global approach offers the benefit of capturing electronic shell structure over large regions of parameter space, building highly-resolved tables in minutes on a simple desktop. We present shock Hugoniot and off-Hugoniot calculations for a number of materials, comparing \SpK~to other models and experimental data. We also apply EoS and opacity data generated by \SpK~in integrated simulations of indirectly-driven capsule implosions, highlighting physical sensitivities to the choice of EoS models.
\end{abstract}
\newline\newline
\textbf{Keywords:} Global equation of state data, High-energy-density matter, Inertial confinement fusion, Electron shell structure, Diatomic molecular dissociation, Maxwell construction
\vspace{1cm}]
\section{Introduction}
\label{section:intro}
\footnotetext{* Corresponding author: adam.fraser@firstlightfusion.com}
The application of radiation-magnetohydrodynamics (RMHD) codes to the modelling of high-energy-density physics experiments is critical to the understanding of the systems being modelled and for informing the direction of experimental campaigns \cite{HatfieldNature2021, WeberPoP2017, WeberPoP2020, WalshPoP2021, ChittendenPoP2016, AmendtPoP2019}. The system of RMHD equations governing the evolution of physical systems must be closed by supplying information about the microphysical properties of matter. For instance, the equation of state (EoS) gives the relationships between thermodynamic quantities, e.g., pressure, internal energy, mass density and temperature. Additional closures are required to describe transport and relaxation processes, such as 
the fluxes associated with conduction and radiation, viscous stresses, particle diffusion, chemical/nuclear reactions, drag between flows and energy exchange in multi-temperature systems \cite{Larsen2017, Drake2018, DeBortoli_book}. These properties are emergent from the microphysical system; e.g., collisions, long-range correlations and interactions with dynamic electromagnetic fields, and are derived from suitable kinetic theories of matter \cite{Lifshitz_book, Bonitz_book, Simakov_PhysPlasmas_2016}. Simple parametric fits and scaling laws \cite{NRL_formulary} are often used for computational efficiency, being justifiable if simulations are initialised under suitable conditions or evolve through regimes where such simple closures are inappropriate so rapidly that their impact on the states of interest can be neglected. However, integrated modelling and design work of ICF targets and supporting experiments generally require detailed knowledge of the evolution of matter across phase boundaries and into states characterised by non-ideal thermodynamic properties \cite{KrempSpringer2006}. 

The requirement of spanning multiple orders of magnitude in mass density and temperature, multiple materials across different initial and final phases and over a broad range of physical phenomena places a substantial burden on models designed to supply closure data to RMHD codes. Start-of-the-art methods based on ab initio simulations \cite{Gonze_CompMatScience_2002} are generally very computationally expensive, with only low-resolution tabulated global data available for a limited number of materials \cite{MilitzerPRE2021}. Such high-fidelity methods are invariably based on the \lq{physical picture}\rq, in which no distinction is made between bound and free electrons, and all emergent thermodynamics arise from fundamental relationships and structural properties. Transport coefficients can be calculated in the local limit from autocorrelation functions between observable quantities (particle density, velocity, energy, etc.) \cite{GreenJChemPhys1954, KuboJPhysSocJpn1957, Greenwood_ProcPhysSoc_1958, ScheinerPRE2019, ChengPRL2020}. Despite the tremendous power and rapid rate of progress of ab initio capabilities \cite{HuPRE2017, BethkenhagenPRR2020, GaoPRB2023}, however, there remains a paucity of fully self-consistent closure data for all but a few select materials and, even then, severe restrictions exist on the range of conditions for which high-fidelity results are available. Thus, it remains of substantial interest to develop utilitarian, lower-fidelity capabilities which can be deployed universally, requiring far more modest computational capacity and iterated rapidly in uncertainty quantification studies.

A simpler approach is to appeal to the 
\lq{chemical picture}\rq, in which electrons bound to ions of different charges are considered distinct chemical species with their own partition functions. Within this description the methods of statistical mechanics can be used to calculate the desired properties of the system. The partition functions can be constructed provided the energies of the bound states are known. For global models, this information is often provided from the calculated energy levels of the isolated atom \cite{MacFarlaneHEDP2007, KilcreaseHEDP2015, ZaghloulHEDP2018}. This approach becomes inaccurate at high densities, when interparticle coupling is high and the bound states become significantly perturbed from their isolated-atom energies due to correlations with other charges in the surroundings and degeneracy effects \cite{HuPRL2017}.

Another approach is based on \lq{average-atom}\rq~calculations, where a single atom representative of the ensemble average of all others in the system is modelled. The finite-temperature Thomas-Fermi (TF) model \cite{FeynmanPR1949} is a simple example of such a model that modern EoS codes such as the \texttt{FEOS} package \cite{FaikCPC2018} still used today. The thermodynamic properties of the TF model obey scaling laws with the atomic and mass number, meaning a single table can be interpolated on for all elements. Some shortcomings of the model include a significant degree of ionisation of low temperature gases when they should be neutral and the lack of atomic shell structure, which leads to the most significant error in low atomic number (low Z) elements. This results in EoS data for the hydrogen isotopes, of particular importance to fusion research, being particularly inaccurate. The screened hydrogenic model with $\ell$-splitting (SHM-$\ell$), developed by Faussurier et al.~\cite{FaussurierHEDP2008}, is an average atom calculation that accounts for atomic shell structure. However, Faussurier et al.~only apply the model above temperatures of a few eV, below which they interpolate into the TF model. This negates their approach as an option for improving upon the low-temperature shortcomings of the TF model.

The atomic and microphysics code, \SpK~\cite{CrillyHEDP2023}, works in the chemical picture. It solves the Saha equation, modified to account for perturbative, non-ideal effects, to obtain distributions of ion charges and level populations for given conditions. The energy levels of bound states are obtained from the NIST database or calculated using the SHM-$\ell$ where these are unavailable. From the ionisation equilibrium, it calculates the opacity. It also contains the non-local thermodynamic equilibrium (NLTE) \lq{effective temperature}\rq~model of Busquet et al.~\cite{BusquetPRA1982, BusquetPOFB1993, BusquetJQSRT2006}. Details of the atomic, opacity, and NLTE models can found in Ref.~\cite{CrillyHEDP2023}. 

In this work, we present extensions to the existing capabilities of \SpK~to enable the generation of tabular EoS data. We also interpolate into the TF model, but only at high-coupling where the modified Saha approach has greater EoS error. By doing so, we capture the neutral gas phase and electron shell structure in weakly coupled plasma states, enabling the fast computation of global EoS data that is valid over a larger range of conditions than chemical picture and TF models alone.

In Section \ref{section:EoS_components}, we outline the different components of the EoS model, including an approximation of the physics of diatomic molecular dissociation. Next, Section \ref{section:Hugoniot} applies EoS data generated by \SpK~to shock Hugoniot calculations, benchmarking against other models and validating against experimental data. In Section \ref{sec:off_hugoniot}, we compare the output of \SpK~against ab initio data and other models in examples of off-Hugoniot states. Then, Section \ref{section:ICF} applies \SpK~data in integrated simulations of 1-dimensional (1D) indirectly-driven ICF implosions, demonstrating and explaining sensitivities of results to the EoS models used. Finally, Section \ref{section:discussion} discusses the outcomes of the work and presents avenues for further improvement of \SpK.

\newcommand{\kC}{k_{\mathrm{C}}}
\newcommand{\kB}{k_{\mathrm{B}}}
\newcommand{\kT}[1]{\kB T_{#1}}

\section{Equation of State components} \label{section:EoS_components}

To generate global EoS tables, \SpK~follows the commonly used method of free energy decomposition, whereby contributions due to different species in the plasma are treated as distinct and decoupled \cite{FaikCPC2018, GaffneyHEDP2018}. The Helmholtz free energy (denoted from here on as simply the \lq{free energy}\rq~unless otherwise stated) of the total system is thus approximated as a sum of three terms:
\begin{equation}
    \label{eq:free_energy_decomposition}
    F(\rho, T_{e}, T_{i})
    = 
    F_{e}(\rho, T_{e}) + F_{i}(\rho, T_{i}) + F_{b}(\rho, T_{e})
    \,.
\end{equation}
Here, $F_{e}$ represents the contribution from the electrons, $F_{i}$ from the ions, and $F_{b}$ is the term introduced to accommodate chemical bonding \cite{More1986} and is applied as a bulk correction, irrespective of the detailed chemical composition of a particular material; it will be described in greater detail later. 

From Eq.~\eqref{eq:free_energy_decomposition}, the principal thermodynamic quantities of interest to RMHD codes; the partial pressures $p_{\alpha}$, entropies $S_{\alpha}$ and internal energies $U_{\alpha}$, can be derived using Maxwell's relations:
\begin{align}
    \label{eq:pressure_def}
    p_{\alpha}
    = &\,
    -\left.\frac{\partial F_{\alpha}}{\partial V}\right|_{N_{\alpha}, T_{\alpha}}
    \,,
    \\
    \label{eq:entropy_density_def}
    S_{\alpha}
    = &\,
    -\left.\frac{\partial F_{\alpha}}{\partial T_{\alpha}}\right|_{N_{\alpha}, V}
    \,,
    \\
    \label{eq:internal_energy_density_def}
    U_{\alpha}
    = &\,
    F_{\alpha} + T_{\alpha}S_{\alpha}
    \,,
\end{align}
for components $\alpha \in \{e, i, b\}$. Note that in the first two terms of Eq.~\eqref{eq:free_energy_decomposition} the temperature dependence is strictly only due to the particular species, i.e., the free energy of the electrons depends only on the electron temperature, whereas the ionic part depends only on the ion temperature. The bonding term is chosen to be influenced only by the free-electron temperature \eqref{eq:free_energy_decomposition} to remain consistent with the QEOS framework. This allows the amalgamation of the free-electron and bonding terms into a single component of the EoS tables produced by the code. This assumption is reasonable for fully ionised plasmas, but can be erroneous in strongly two-temperature states featuring partial ionisation; this deficiency is also noted by More et al., but remains prevalent in QEOS-type models, e.g., \texttt{FEOS} \cite{FaikCPC2018}.

The extension to the \SpK~code reported here represents a substantial improvement over models such as \texttt{FEOS} by including electronic shell structure in the weakly coupled limit and allows a sufficiently accurate approximation of degenerate and non-plasma phases, many or all of which are encountered in integrated simulations of HEDP systems. This section will outline the various models that comprise the global \SpK~EoS. Beginning with the electron model, the Q-MHD \cite{NayfonovApJ1999} (Section \ref{section:QMHD}) and TF \cite{MorePhysF1988} (Section \ref{section:TF_model}) models respectively detail the low- and high-coupling behaviour of the electron EoS. The means by which they are interpolated between is then outlined in Section \ref{section:interpolation}. The Cowan model \cite{MorePhysF1988} is utilised for the ion EoS. Modifications implemented in \SpK~to the basic model that improve its robustness or capture additional physics are then outlined in Section \ref{section:Cowan}. The bonding corrections used to correctly capture material properties under ambient conditions are described in Section \ref{section:bonding}, with the Maxwell construction implemented to capture the liquid-vapour coexistence region detailed in Section \ref{section:Maxwell}.

\subsection{The Q-MHD electron EoS}
\label{section:QMHD}

Initially developed as an opacity code, \SpK~predicts distributions of ion charges and level populations based on a solution to a modified Saha equation \cite{CrillyHEDP2023}. For weakly coupled plasmas, \SpK~utilises the electronic component of the Q-MHD model \cite{NayfonovApJ1999}. The electronic Helmholtz free energy is decomposed into contributions from the free electrons, bound electrons, and (non-ideal) Coulombic interaction terms, i.e. 
\begin{align}
    \label{eq:free_energy_QMHD}
    F_{e}(\rho, T_{e})
    = &\,
    -N_{e}\kT{e}
    \left[\frac{2V}{N_{e}\Lambda_{e}^{3}}
    \mathcal{F}_{3/2}(\eta_{e})
    - \eta_{e}
    \right]
    \nonumber\\ 
    &+
    F_{e}^{\mathrm{int}}(\rho, T_{e})
    +
    F_{e}^{\mathrm{IPD}}(\rho, T_{e})
    \,.
\end{align}
Note that we do not currently account for an exchange-correlation contribution in the electron gas \cite{GrothPRL2017} since the Q-MHD model is used only under conditions where strong coupling and exchange effects are small (see Section \ref{section:interpolation}). An extended treatment including this missing physics is under development and will be reported in future work.

In Eq.~\eqref{eq:free_energy_QMHD}, the number of free electrons is related to the mass density and average ionisation state of the system (computed from the atomic model of \SpK) as $N_{e} = \sum_{\alpha}Z_{\alpha}N_{\alpha} = V \left\langle Z \right\rangle \rho / \left\langle m \right\rangle$, where $\left\langle Z\right\rangle = \left(\sum_{\alpha}Z_{\alpha}N_{\alpha}\right)/\sum_{\alpha}N_{\alpha}$ and $\left\langle m\right\rangle = \left(\sum_{\alpha}m_{\alpha}N_{\alpha}\right)/\sum_{\alpha}N_{\alpha}$; $Z_{\alpha}$ and $m_{\alpha}$ are the charge and mass of an ion of species $\alpha$, respectively. Furthermore, the 
thermal de Broglie wavelength is $\Lambda_{e} = h/(2\pi m_{e}\kT{e})^{1/2}$ and $\eta_{e} = \mu_{e}/\kT{e} = \mathcal{F}_{1/2}^{-1}(N_{e}\Lambda_{e}^{3}/2V)$ is the dimensionless ideal chemical potential of the free electrons. The function $\mathcal{F}_{j}(\eta)$ is the complete Fermi-Dirac integral of order $j$ \cite{KrempSpringer2006}; these (and their inverses) are implemented using the rapid and accurate rational approximations given by Fukushima \cite{FukushimaAMC2015}.

The first term of Eq.~\eqref{eq:free_energy_QMHD} represents the translational component, i.e., due to the three-dimensional motions of the electrons; it can be readily verified to reduce to the classical expression in the limit $\eta_{e} \to -\infty$ (see, e.g., Ref.~\cite{Kahlbaum_FluidPhaseEquil_1992}). The second term refers to the internal partition function and contains the ionisation and excitation energies of the bound states. This is intimately connected to the ionisation equilibrium predicted by the DCA part of \SpK, as detailed in Ref.~\cite{CrillyHEDP2023}, from which it takes all its data. The last term gives the contribution arising from the (screened) Coulomb interactions between the free electrons and the ions. This term is responsible for introducing the notion of ionisation potential depression (IPD) \cite{CiricostaPRL2012, HoartyPRL2013, KrausPRE2016}. We presently limit ourselves to the simple ion sphere (IS) and Debye-H\"{u}ckel (DH) models for simplicity, despite more general and accurate models being available \cite{Crowley2014, LinPRE2017}.

In the Q-MHD model, the internal partition function that models the bound electrons is truncated by reducing the statistical weights of bound states using smooth functions that approximate the effects of the oscillating plasma microfield and delocalisation driven by the close proximity of particles at high densities. The details of the internal model can be found in Ref.~\cite{NayfonovApJ1999}.

\begin{figure}[!ht]
    \centering
    \includegraphics[width=0.45\textwidth]
    {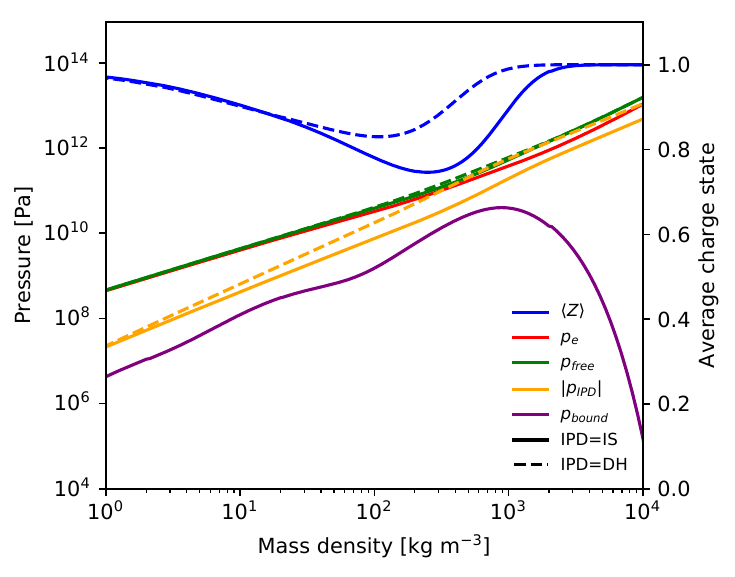}
    \caption{
        The contributions to the total electron pressure (red) in deuterium made by the free electrons (green), the negative-valued IPD term (orange), and the bound electrons (purple) along an isotherm at 10 eV. The IS (solid) and DH (dashed) IPD models are compared showing the impact on the pressure and the average charge state (blue).
    }
    \label{fig:D2_pressure_10eV}
\end{figure}

The various contributions to the electron pressure are shown in Fig.~\ref{fig:D2_pressure_10eV} for a deuterium plasma along a 10 eV isotherm, alongside the average charge state. The impact of the thermodynamic contribution to the pressure made by the IPD term is negative, so its absolute magnitude is plotted. The influence of the IPD is evident in both the ionisation and the associated contributions to the total pressure. The role of this crucial aspect of the physics is substantially increased at lower temperatures, where the differences between models becomes even more stark. Practically speaking, however, such differences are of little consequence in \SpK, since they are subsumed within the interpolation to the TF model (see later discussion).

\subsection{Multi-species Thomas-Fermi electron EoS}
\label{section:TF_model}

Whilst the detailed balance of free and bound states in the chemical picture can be included with the Q-MHD approach, it is only as accurate as the information provided by ionisation equilibrium derived from the modified Saha-Boltzmann equation. Unfortunately, it is well-known that Saha-type models become inaccurate as the degree of interparticle coupling in a plasma increases. A more appropriate model for dense plasmas, albeit at the expense of the detailed insight to the ionisation equilibrium, is the well-known finite-temperature Thomas-Fermi (TF) average atom model \cite{FeynmanPR1949, MorePhysF1988}. However, rather than appealing to a straightforward average atom picture, we have implemented the mixture scheme following the suggestion of More et al.~\cite{MorePhysF1988}. This is realised in \SpK~with an iterative scheme for obtaining values for the average charge state of each element (and, thus, also the overall average) in a plasma composed of an arbitrary number of elemental components. The approach modifies the average volume occupied by the average ion of each species, conserving the total volume, until each predicts a consistent electron density at the boundaries of their respective ion spheres. This is the appropriate approach for the TF model as it ensures no electric fields at ion sphere boundaries; a boundary condition of the TF model. 

The fit by More \cite{Atzeni2009} is used for a fast estimate of the partial volumes of each element. This partial volume is then used to interpolate on a table of solutions to the TF model that includes thermodynamic variables of interest, scalable to the element in question using the convenient laws \cite{MorePhysF1988}. The partial contributions due to each element to the total thermodynamic variables are then calculated using
\begin{equation}
    X 
    = 
    \sum_{s}x_{s}X_{s}
    \,,
\end{equation}
where $X_{s}$ is the partial contribution to the total thermodynamic quantity, $X$, made by species $s$, and $x_{s}=n_{s}/n_{i}$ is the fraction the species makes up of the total ion density, denoted by $n_{i}$.

\subsection{Defining an interpolation locus between the Q-MHD and TF models}
\label{section:interpolation}

In the interest of preserving shell structure within the EoS for as much of parameter space as possible, we utilise the modified Saha component of the EoS up to the boundary of its validity. \SpK~possesses the capability to choose from multiple measures that can be used to define a locus of points that set the boundary between the TF and modified Saha regions of the table. For instance, a screening- and degeneracy-corrected ion-ion coupling parameter can be written \cite{MurilloPRE2013, GerickePRE2002},
\begin{equation} 
    \label{eq:SpK_interp_ii_screen}
    \tilde{\Gamma}_{ii}
    =
    \frac{\bar{Z}^{2}e^{2}}{4\pi\epsilon_{0} R_{\text{WS}}\langle E_{kin}\rangle}
    \exp\left(-\frac{R_{\text{WS}}}{\lambda_{s}}\right)
    \,.
\end{equation}
Here, $\bar{Z}$ is the average ion charge in the plasma, $e$ is the electron charge, $\epsilon_{0}$ is the permittivity of free space, $R_{\text{WS}} = (4\pi n_{i}/3)^{-1/3}$ is the Wigner-Seitz radius, $\lambda_{s}$ is the plasma screening length, and $\langle E_{kin}\rangle$ is the degeneracy-corrected average kinetic energy \cite{GerickePRE2002},
\begin{equation}
    \langle E_{kin}\rangle 
    =
    \frac{3k_{\text{B}}T_{e}}{n_{e}\Lambda_{e}^{3}}\mathcal{F}_{3/2}(\eta_{e})
    \,.
\end{equation}

\begin{figure}[!ht]
    \centering
    \includegraphics[width=0.45\textwidth]{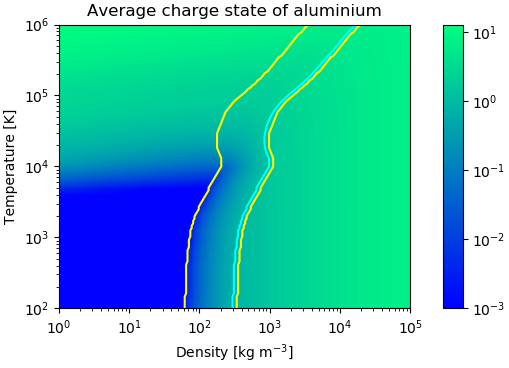}
    \caption{
        The average charge state of aluminium for a range of densities and temperatures. The yellow curves show the boundaries of the interpolation between the modified Saha model on the left and the TF model on the right. The cyan curve shows where the ion-ion coupling parameter is unity.
    }
    \label{fig:Al_interpolation}
\end{figure}

With the boundary between the two regions of the EoS table set by a chosen parameter value, a user-defined range in the chosen parameter or relative density range is used to set a window between which the two models are interpolated. The present interpolation method simply uses cubic spline interpolation in the thermodynamic variables of interest. An example of this is shown in Fig.~\ref{fig:Al_interpolation}. Whilst this is not ideal, as it introduces a degree of thermodynamic inconsistency, the impact of such a deficiency is generally of extremely limited consequence for simulations. We are presently developing an improved method, which will be reported in future work.

\subsection{Modifications to the Cowan model}
\label{section:Cowan}

The Cowan model is well-known for its use as the ionic component of the QEOS model \cite{MorePhysF1988}. It is fast and generalised, requiring only a material's average atomic and mass numbers for initialisation. Some improvements and additional physics included in the model are detailed below.

\subsubsection{Smoother transitions between model phases}

The Cowan model has 3 distinct phases: the fluid, the hot solid, and the cold solid. Which phase the model lies in depends on the ratio between the ion temperature, $T_{i}$, and the density-dependent melting, $T_{m}(\rho)$, and Debye, $\Theta_{D}(\rho)$, temperatures. That is, the fluid phase is found where $T_{i}>T_{m}$, and the hot solid is found where $3T_{i}>\Theta_{D}$. More et al.~suggest the transition is nearly smooth. However, the discontinuities transitioning between the models can sometimes be significant enough to cause problems with the Maxwell construction.  
Two modifications to the structural model are presented here: one that interpolates the pressure between the solid and liquid models, instead of a hard switch at the boundary between the two; another that replaces the transition between the hot and cold solid models with a numerical integration of the ideal phonon energy spectrum in the Debye model \cite{BlundellOUP2010}. 

In the case of the solid-liquid transition, a sigmoid function is utilised to define the fraction of material in either the solid or liquid state, e.g., with respect to the ion pressure:
\begin{align}
    f_{\mathrm{liq}} 
    = &\,
    \left[1 + e^{-\tau/\tau_{0}}\right]^{-1}, ~~~ \tau = 
    \frac{T_{i}}{T_{\mathrm{melt}}} - \frac{T_{\mathrm{melt}}}{T_{i}}
    \,, 
    \\
    p_{i} 
    = &\,
    f_{\mathrm{liq}} p_{\mathrm{liq}} 
    + 
    (1 - f_{\mathrm{liq}}) p_{\mathrm{sol}}
    \,.
\end{align}
Here, $T_{\mathrm{melt}}$ is the (density-dependent) melting temperature, $\tau_{0}$ is a user-defined scale over which the interpolation acts, $p_{i}$ is the total ion pressure, $p_{\mathrm{liq}}$ is the liquid pressure, $p_{\mathrm{sol}}$ is the solid pressure for the same temperature, and $f_{\mathrm{liq}}$ is the fluid phase fraction.

For the solid model, the free energy per atom is described by averaging the phonon energies across the frequency distribution,
\begin{equation} 
    \label{eq:F_solid_Cowan}
    F_{at}(\rho, T_{i}) 
    = 
    \int_{0}^{\nu_{\text{D}}}
    \text{d}\nu\,
    g(\nu)\,F(\nu, T_{i})
    \,,
\end{equation}
where $\nu_{\text{D}}(\rho)$ is the Debye frequency (known via the Empirical model), $g(\nu)=9\nu^{2}/\nu_{\text{D}}^{3}$ is the density of frequency states per atom, and $F(\nu, T_{i})$ is the free energy of a given normal mode,
\begin{equation}
    F(\nu, T_{i}) 
    = 
    \frac{h\nu}{2} + k_{\text{B}}T_{i}\ln\left(1 - e^{-h\nu/k_{\text{B}}T_{i}}\right)
    \,.
\end{equation}
The expressions for the Helmholtz free energy of the low- and high- temperature solids in More et al.~\cite{MorePhysF1988} are obtained by expanding the log term in the limiting cases of $u=h\nu/k_{\text{B}}T_{i}$ being small or large and integrating by parts. 

In \SpK, the cold- and hot-solid models are replaced by a single solid model that numerically integrates Eq.~\ref{eq:F_solid_Cowan} by adaptive 8-16 Gauss-Legendre quadrature. The cold solid model is only used when the temperature is so low that the accuracy gained by the more expensive numerical integration is negligible.

\subsubsection{Approximate diatomic molecular dissociation model}

A molecular substance that is sufficiently heated will undergo molecular dissociation into individual atoms or ions. Though this is a chemical reaction, the result of the process is to change the number of particles in the system that are more appropriately described by the ion temperature. Therefore, within the current framework the thermodynamic impact of molecular dissociation is reflected in the ion EoS. 

Given the importance of substances such as deuterium ($\ce{D_2}$) and tritium ($\ce{T_2}$) in fusion systems, and the considerable complexity of a more generalised treatment for polyatomic gases and monomeric/polymeric materials, we limit ourselves here to only diatomic molecules. \SpK~utilises a simple model for diatomic molecular dissociation. Consider a mixture of dissociated atoms, $A$, and the diatomic molecules they can form, $A_{2}$. The atoms that make up the molecule could be nonidentical, though the Cowan model assumes a single species with atomic and mass numbers equivalent to the concentration-average over the species in a material. In the chemical picture, the equilibrium between molecules and dissociated atoms is found via the relation \cite{YoungJAP1995, Zeldovich1966, BlundellOUP2010},
\begin{equation} 
    \label{eq:molecular_equilibrium}
    K 
    = 
    \frac{N_{A}^{2}}{N_{A_{2}}} 
    = 
    \frac{Q_{A}^{2}}{Q_{A_{2}}}
    \,,
\end{equation}
where $N_{X}$ and $Q_{X}$ are the respective number of particles and the canonical partition function of species $X=A_{2},A$. A common approximation is the decomposition of the molecular partition function into the product of decoupled modes \cite{Zeldovich1966, BlundellOUP2010},
\begin{equation}
    \label{eq:molecular_partition_function}
    Q_{A_{2}} 
    = 
    Q_{A_{2},\text{trans}}
    Q_{A_{2},\text{int}}
    Q_{\text{vib}}
    Q_{\text{rot}}
    \,,
\end{equation}
where $Q_{A_{2},\text{trans}}$, $Q_{A_{2},\text{int}}$, $Q_{\text{vib}}$, and $Q_{\text{rot}}$ are the translational, internal, vibrational, and rotational partition functions, respectively. The rotational partition function is usually expressed by a quantised sum over rotational energy states. For high temperature applications, the summation can be replaced by an integral with the result \cite{Zeldovich1966},
\begin{equation}
    Q_{\text{rot}} 
    = 
    \frac{\kT{i}}{B_{m}}\frac{1}{\sigma}
    \,,
\end{equation}
where $B_{m}=\hbar^{2}/2I_{m}$ is the rotational constant, set by the moment of inertia, $I_{m}$. The symmetry factor takes a value of $\sigma=1,2$ for diatomic molecules with different or identical nuclei, respectively. The vibrational partition function is expressed by \cite{Zeldovich1966},
\begin{equation} 
    \label{eq:Q_vib}
    Q_{\text{vib}} 
    = 
    \left[1 - \exp\left(-\frac{\hbar\omega_{m}}{\kT{i}}\right)\right]^{-1}
    \,,
\end{equation}
where $\hbar\omega_{m}$ is the vibrational constant. The factor corresponding to the zero-point energy, $\exp(\hbar\omega_{m}/2\kT{i})$ is not present in Eq.~\ref{eq:Q_vib} because it can be absorbed into the ground state energy of the internal component of the molecular partition function. Note that throughout the foregoing, the temperature appropriate to the discussion of the molecular contribution is \textit{always} that of the ions.

Rather than requiring the full internal partition functions of both the atomic and molecular species, the fact that the most populated state in both species will be the ground state implies \cite{Zeldovich1966},
\begin{equation}
    \frac{Q_{A,\text{int}}^{2}}{Q_{A_{2},\text{int}}}
    \simeq  
    \frac{g_{A,0}^{2}}{g_{A_{2},0}}
    \exp\left(-\frac{E_{D}}{\kT{i}}\right)
    \,.
\end{equation}
Here, $g_{X,0}$ is the statistical weight of the ground state of species $X \in \{A_{2}, A\}$, and $E_{D} = 2E_{A,0} - E_{A_{2},0}$ is dissociation energy of one molecule, such that the ground state atomic configuration energy, $E_{A,0}$, will lie at an energy $E_{D}/2$ above those of molecules $E_{A_{2},0}$. The dissociation energy, vibrational constant, and rotational constant are measured quantities and are available for a number of materials on the NIST database.

For the case of diatomic molecular dissociation and assuming only the molecular and dissociated neutral atomic states can exist (i.e. ignoring the possible ionic states), the concentration of each species can be obtained by solving the quadratic,
\begin{equation}
    N_{A}^{2} +\frac{K}{2}N_{A} - \frac{K}{2}N_{\text{tot}} 
    = 0
    \,,
\end{equation}
where $K$ is defined in Eq.~\ref{eq:molecular_equilibrium} and $N_{\text{tot}} = N_{A}+2N_{A_{2}}$ is the total number of nuclei. 

Given the simplified forms used, at high temperatures the rotational and vibrational partition functions can eventually become large enough to erroneously lead to recombination. We use a cut-off function to truncate the molecular partition function at high temperatures, replacing $Q_{A_{2}}$ in Eq.~\ref{eq:molecular_equilibrium} with $Q_{A_{2}}^{*}=Q_{A_{2}}Q_{\text{cutoff}}$. The cut-off function is defined as,
\begin{align}
    Q_{\text{cutoff}}
    = &\,
    2\left[q_{+}(T_{i})\left(\frac{T_{i}}{T_{\text{max}}}\right)^{2} + q_{-}(T_{i})\right]^{-1}
    \,,
    \\
    q_{\pm} 
    = &\,
    1 \pm \text{erf}(T_{i} - T_{\text{max}})
    \,,
\end{align}
where $\text{erf}(x)$ is the error function and $\kB T_{\text{max}} = E_{D}$.

The model should ensure molecules undergo pressure dissociation at sufficient densities. We use a combination of the density-dependent dissociation energy of Young and Corey \cite{YoungJAP1995} and include an \lq{effective temperature}\rq, 
\begin{equation}
    T^* 
    = 
    T_{i} + C\frac{\rho}{\rho_{s}}
    \,,
\end{equation}
where $\rho_{s}$ is the mass density at the reference conditions and $C$ is a user-defined constant with units of K. We note that a configurational partition function that represents the interaction between molecules should technically be included in the model \cite{RossPRB1998, BatesCPC2003}, which would have most impact at higher densities and could replace the need for this effective temperature.

Accounting for molecular equilibrium in the Cowan model requires the Helmholtz free energy to be the sum of that of both the atoms and molecules. If $\alpha_{D}=N_{A}/N_{\text{tot}}$ is the fraction of the molecules that are dissociated, the free energy is then
\begin{align} 
    \label{eq:molecule_free_energy}
    F 
    = &\,
    -\biggl\{
        \frac{\alpha_{D}E_{D}}{2} 
        +
        \kT{i}
        \biggl[
            \alpha_{D}\ln(Q_{A})
            +
            \frac{1 - \alpha_{D}}{2}
            \ln(Q_{A_{2}})
        \biggr]
    \biggr\}
    \,.
\end{align}
Given the scope and simplicity of Young and Corey's model, the atomic partition function contains only a translational contribution, i.e., $Q_{A} = Q_{A,\text{trans}}$, neglecting any correlation effects. Similarly, only the gas-like terms: translation, rotation and vibration, are included in the molecular partition function as per Eq.\;\eqref{eq:molecular_partition_function}. Inclusion of entropy of mixing and suitable correlation contributions are presently under development and will be reported on in future work. The translational contributions in Eq.\;\eqref{eq:molecule_free_energy} are given by the standard expressions in the Cowan model and the contribution by the internal partition functions is accounted for by the final term.
All other thermodynamic quantities are then obtained through the appropriate thermodynamical derivatives of Eq.~\ref{eq:molecule_free_energy}.

\subsection{Bonding corrections}
\label{section:bonding}
 
One of the most useful assumptions in average atom models is the spherical symmetry of the potential surrounding any particular ion in the system. This helps to reduce the complexity of the mathematical framework and is appropriate across the plasma phase of the EoS. However, it also directly leads to the shortcoming that chemical bonding is not emergent. It is well-known that the TF model predicts a substantial electron pressure in conditions where one should have a solid at atmospheric pressure \cite{MorePhysF1988}. In the QEOS model and codes such as \texttt{FEOS} that are derived from it, this problem is remedied by the use of \textit{ad hoc} bonding corrections \cite{FaikCPC2018, MorePhysF1988}. 

 The bonding corrections utilised in \SpK~are identical to those in \texttt{FEOS}: the semi-empirical Barnes correction \cite{Barnes1967}, and the soft-sphere correction introduced by Young and Corey \cite{YoungJAP1995}. In the Barnes correction, a Morse-like potential is used to describe the pressure behaviour along the cold curve. The repulsive term in the potential is well-approximated by the TF model and the attractive term is defined such that the model predicts zero pressure at solid density along the cold curve with the correct bulk modulus. The equivalent energy term is calculated thermodynamically consistently through integration. By ensuring the correct bulk modulus for materials, this Barnes correction produces good results when considering compression from the solid state.

The soft-sphere correction replaces the cold curve below solid density. With a supplied cohesive energy, it ensures the correct energy is reached in the low-density limit. Additional parameters can also be tuned to improve agreement with of the critical point with experimental data \cite{YoungJAP1995, FaikCPC2018}. It ensures the pressure is continuous across the solid-density boundary between the two models, though the pressure gradient will be discontinuous. Detailed descriptions of the two models can be found in Refs.\cite{FaikCPC2018, Fraser2023Thesis}.
 
\subsection{Maxwell construction for the liquid-vapour coexistence region}
\label{section:Maxwell}

The attractive terms introduced by the bonding corrections generate van der Waals loops along isotherms lying below the critical temperature. A Maxwell construction must be performed to establish the location of the binodal curve along isotherms and the pressure at which the phase transition occurs. Whilst metastable states can exist for finite periods of time that do not live on the Maxwell construction, this work does not explore the existence of such states in integrated simulations and so these are not considered further.

To calculate the Maxwell construction, \SpK~ensures Maxwell's `equal area rule' is satisfied, stating that for a given isotherm the areas bound by the equilibrium pressure and the pressure along the van der Waals loop are equivalent. This implies that the following must hold true:
\begin{equation} 
    \label{eq:Maxwell_condition}
    \int_{V_{l}}^{V_{v}}(p-p_{\text{eq}})\text{d}V = 0,
\end{equation}
where $p$ is the pressure along the van der Waals loop, $p_{\text{eq}}$ is the equilibrium pressure, and $V_{l}$ and $V_{v}$ are the respective volumes of the liquid and vapour states that lie on the binodal curve \cite{BlundellOUP2010}. 

Before performing the Maxwell construction, the isotherms are iterated over with increased temperature until the final isotherm is obtained that contains a van der Waals loop. Next, looping through each isotherm, the following process calculates the Maxwell construction:
\begin{enumerate}
    \item The spinodal curve is found by scanning from the extremes of density to find the densities where the pressure gradient changes sign. This will be slightly off from the actual spinodal curve, since only densities are considered that lie on the table stencil. However, with sufficient resolution this should not be a problem, practically.
    
    \item The spinodal maximum and minimum are used as bounds for a binary search to find the equilibrium pressure.
    
    \item The points that would lie on the binodal curve (were their pressure to match the equilibrium pressure) are found by root finding by bisection between the spinodal curve and the extremes of the table. This is done to find the densities (specific volumes) that form limits of the integral in Eq. \ref{eq:Maxwell_condition}.
    
    \item A numerical integration by Gauss-Legendre quadrature is performed on Eq.~\ref{eq:Maxwell_condition} between the volume limits found in the previous step.
    
    \item The bounds of the binary search are updated based on the sign of the integral.
    
    \item Steps 3-5 are repeated until convergence is reached. 
\end{enumerate}

Once converged, the values of the thermodynamic functions are obtained for each point along the binodal curve. Along the Maxwell construction, these variables are obtained by calculating the fractional mass within the liquid, $x_{l}$, or vapour, $x_{v}$, states and using these to take the linear average. For example, the specific (per unit mass) Helmholtz free energy within the liquid-vapour coexistence region is written,
\begin{equation}
    f 
    = 
    f_{\mathrm{vap}}x_{\mathrm{vap}} + f_{\mathrm{liq}}x_{\mathrm{liq}} = f_{\mathrm{vap}}x_{\mathrm{vap}} + f_{\mathrm{liq}}(1 - x_{\mathrm{vap}})
    \,,
\end{equation}
where $f_{\mathrm{vap}}$ and $f_{\mathrm{liq}}$ are the contributions from the vapour and liquid states, respectively. The quantities must be specific for this approach to be valid. The relevant information already known about the system is the total mass density, $\rho$, the density of the liquid, $\rho_{\mathrm{liq}}$, and vapour, $\rho_{\mathrm{vap}}$, states on the binodal curve and the corresponding values of the thermodynamic variables here. The total mass of both states, $M = M_{\mathrm{liq}} + M_{\mathrm{vap}}$, and the total volume $V = V_{\mathrm{liq}} + V_{\mathrm{vap}}$, are known from the total mass density, $\rho = M/V$. Along an isotherm, the densities of the two states do not change, only the mass fraction contained in each. This means $M_{s} = V_{s}\rho_{s}$, for $s\in\{\mathrm{liq}, \mathrm{vap}\}$, and $\rho_{s}$ is constant. Using this information, it can be shown,
\begin{equation} 
    \label{eq:mass_fraction}
    x_{\mathrm{vap}} 
    = 
    \frac{M_{\mathrm{vap}}}{M_{\mathrm{liq}} + M_{\mathrm{vap}}}
    =
    \frac{1/\rho - 1/\rho_{\mathrm{liq}}}{1/\rho_{\mathrm{vap}} - 1/\rho_{\mathrm{liq}}}
    \,.
\end{equation}

\begin{figure}[!hb]
    \centering
    \includegraphics[width=0.42\textwidth]{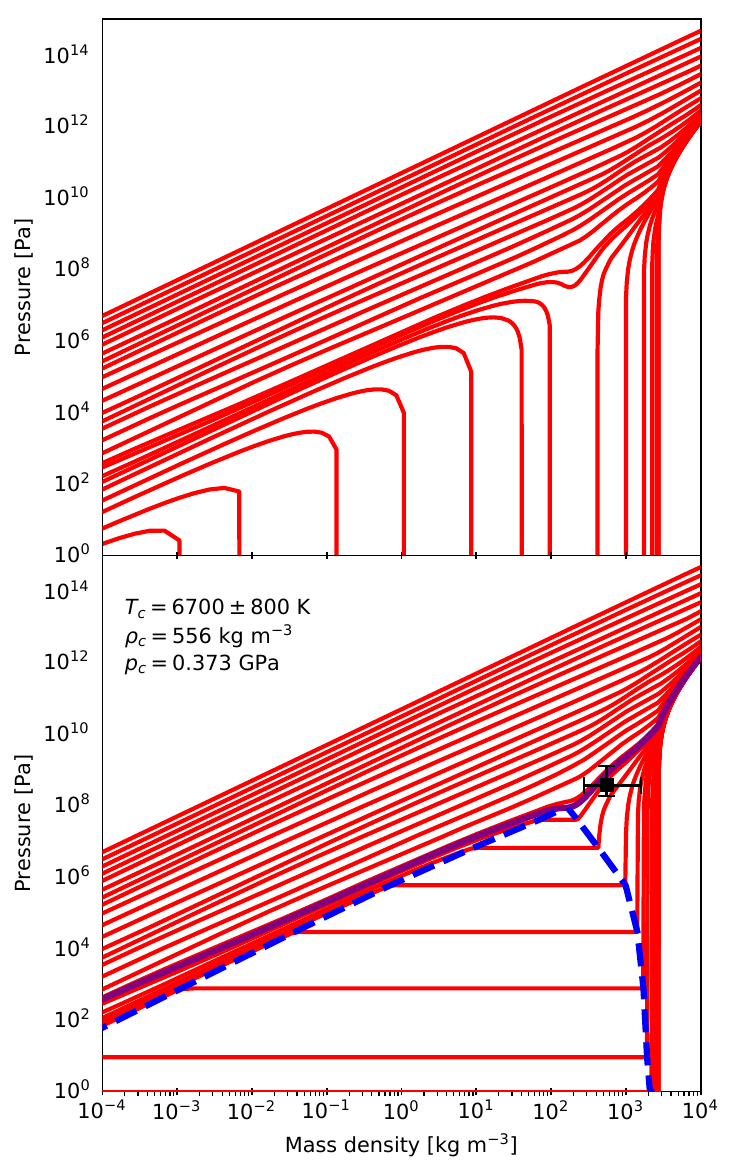}
    \caption{
        The total pressure of aluminium along several isotherms before (top) and after (bottom) a Maxwell construction has been performed. The binodal curve (blue dashed) and the critical isotherm (purple) are also shown. The data for the critical point (black error bars) was taken from Morel et al.~\cite{MorelIJT2009}.
        }
    \label{fig:Al_Maxwell}
\end{figure}

Pressure isotherms for aluminium generated using \SpK~are shown in Fig.~\ref{fig:Al_Maxwell}, comparing the data before and after performing a Maxwell construction. Comparing against the critical point data taken from Morel et al., the estimated critical point data is not yet accurate. Further efforts will include optimisation of free parameters to improve predictions of the critical point. Using the above procedure, \SpK~can be utilised to generate EoS data for use in HEDP simulations that span the full range of conditions expected to occur. Moreover, under weakly coupled conditions the effects of electron shell structure are captured and self-consistent EoS and opacity data can be generated. Thermodynamic consistency is maintained almost everywhere on the table, except for where models are interpolated.

\section{Hugoniot calculations}
\label{section:Hugoniot}

In the following we perform Hugoniot calculations on multiple materials to compare the output of \SpK~against popular global models, highly sophisticated ab initio models, and experimental data. 

\begin{figure}[!ht]
    \centering
    \includegraphics[width=0.45\textwidth]{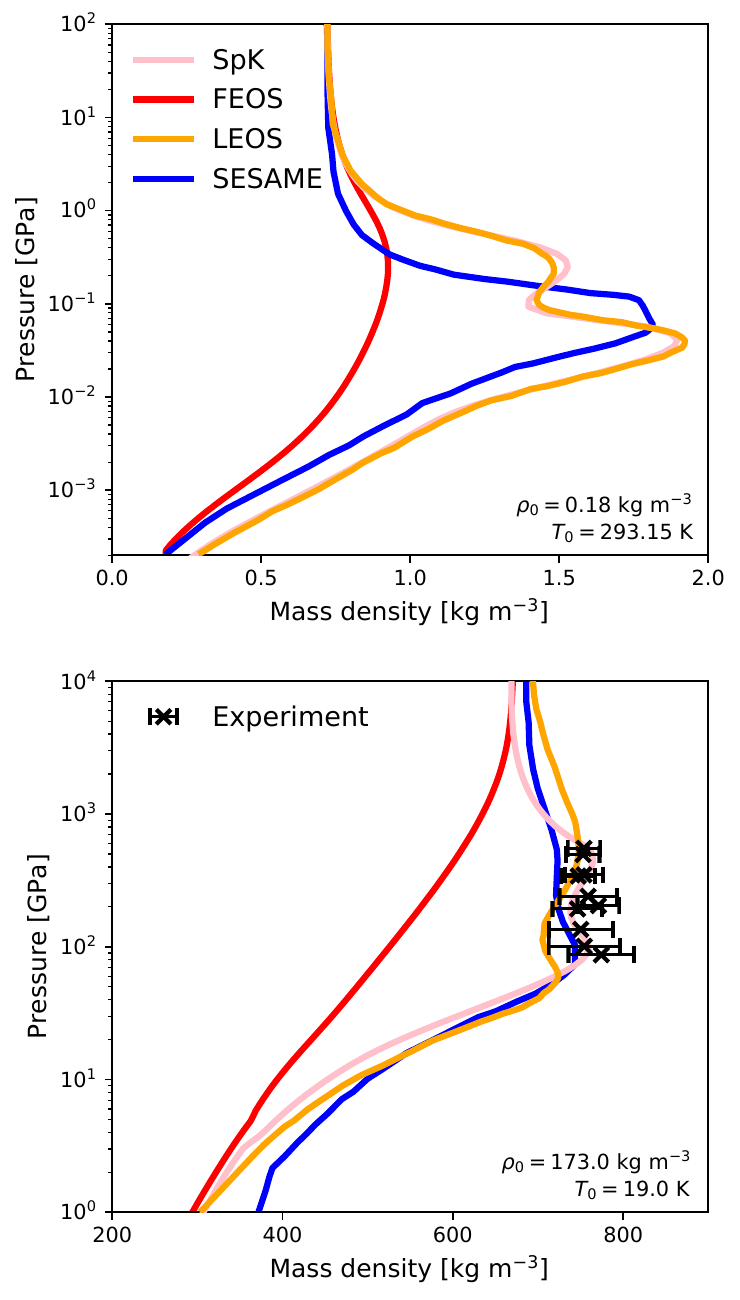}
    \caption{
        Shock Hugoniot for gaseous (top) and cryogenic liquid (bottom) deuterium, with initial conditions shown in each plot. The data shows a comparison between \SpK~(pink), \texttt{FEOS} (red), \texttt{LEOS} (orange), \texttt{SESAME-5263} (blue), and the experimental data of Fernandez-Pa\~{n}ella et al.~\cite{FernandezPRL2019} (crosses).
    }
    \label{fig:D2_Hugoniot}
\end{figure}

The first comparison in Fig.~\ref{fig:D2_Hugoniot} is between shock Hugoniot data for deuterium in either a gaseous initial state at a density of $\rho_{0} = 0.180~\text{kg m}^{-3}$ and temperature of $T_{0} = 293.15$ K, or from a cryogenic liquid at initial density $\rho_{0}=173.0~\text{kg m}^{-3}$ and temperature $T_{0}=19.0 K$. We compare \SpK~against the \texttt{FEOS} package \cite{FaikCPC2018}, \texttt{LEOS} \cite{KerleySNL2003}, and \texttt{SESAME-5263} \cite{Kerley1971, Kerley1972, KerleyPEPI1972}, expecting \texttt{LEOS} to be the most accurate of the latter three models. For the comparison from cryogenic initial conditions, experimental data by Fernandez-Pa\~{n}ella et al.~\cite{FernandezPRL2019} is also included. Two increases in compressibility can be observed along both \SpK~and \texttt{LEOS} Hugoniot curves, corresponding to the deuterium dissociating and then ionising at the lower- and higher-pressure peaks, respectively. The TF model provides electronic component of \texttt{FEOS}, in which the electrons are assumed a semi-classical continuous fluid and lack any shell structure, resulting in ionisation being a process that happens gradually over the full temperature and density range of the table. This is unlike models that contain shell structure, where ionisation occurs in stages above certain temperature thresholds. Thus, neither the sharp increase in compressibility due to both ionisation or dissociation are captured by \texttt{FEOS}. \SpK~is in good agreement with \texttt{LEOS} and is in the best agreement with the experimental data. 

\begin{figure}[!ht]
    \centering
    \includegraphics[width=0.45\textwidth]{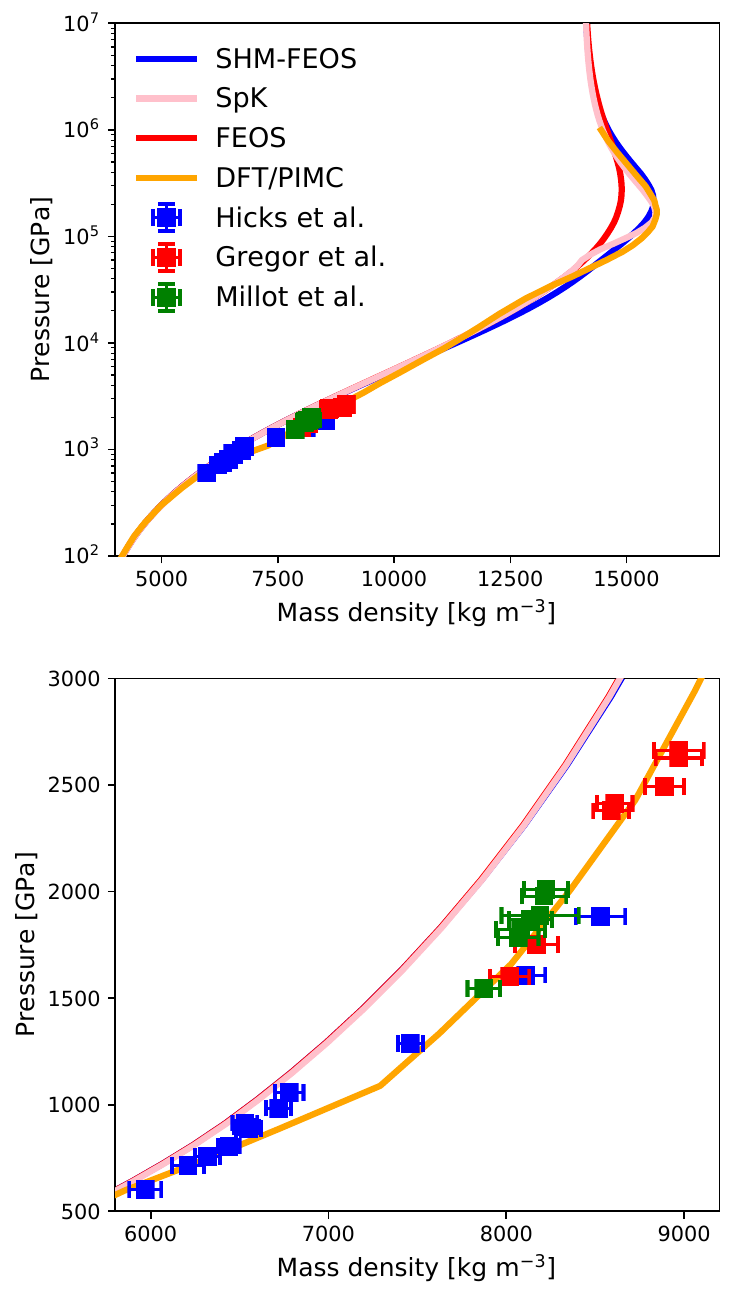}
    \caption{
        Principal Hugoniot for diamond with pressures ranging from $10^{2}$-$10^{7}$ GPa (top) and focusing on the 500-3000 GPa range (bottom). The data shows a model comparison (curves) between \SpK~(pink), the \texttt{SHM-FEOS} model (blue), \texttt{FEOS} (red), the DFT/PIMC data of Benedict et al.~(orange), and the experimental data (squares) of Hicks et al.~\cite{HicksPRB2008} (blue), Gregor et al.~\cite{GregorPRB2017} (red), and Millot et al.~\cite{MillotPoP2020} (green).
    }
    \label{fig:C_Hugoniot}
\end{figure}

Investigating next the principal Hugoniot of diamond, a material also of fundamental importance to ICF experiments \cite{MacKinnonPoP2014, AbuShawarebPRL2022, DavidovitsPoP2022}, Fig.~\ref{fig:C_Hugoniot} shows a comparison between various models and experimental data. Included in the model comparison against \SpK~is the ab initio data of Benedict et al.~\cite{BenedictPRB2014}, \texttt{FEOS} data, and an implementation of the \texttt{SHM-QEOS} model developed by Faussurier et al.~\cite{FaussurierHEDP2008} that replaces the QEOS component with \texttt{FEOS}. Both \SpK~and the \texttt{SHM-FEOS} model utilise the screened hydrogenic model with $\ell$-splitting (SHM-$\ell$) \cite{FaussurierHEDP2008} to incorporate atomic shell structure in calculations, though the former uses a detailed configuration accounting (DCA) approach and the latter is an average-atom model only. There is good agreement between all models except \texttt{FEOS} at the peak in compressibility, due to the K-shell ionisation being effectively captured by the other models. The focus on where the experimental data lies reveals a deviation of the ab initio and experimental data from the other models, due to the high-pressure melting of diamond in these conditions, an effect not yet captured by the simpler models.

\begin{figure}[!t]
    \centering
    \includegraphics[width=0.49\textwidth]{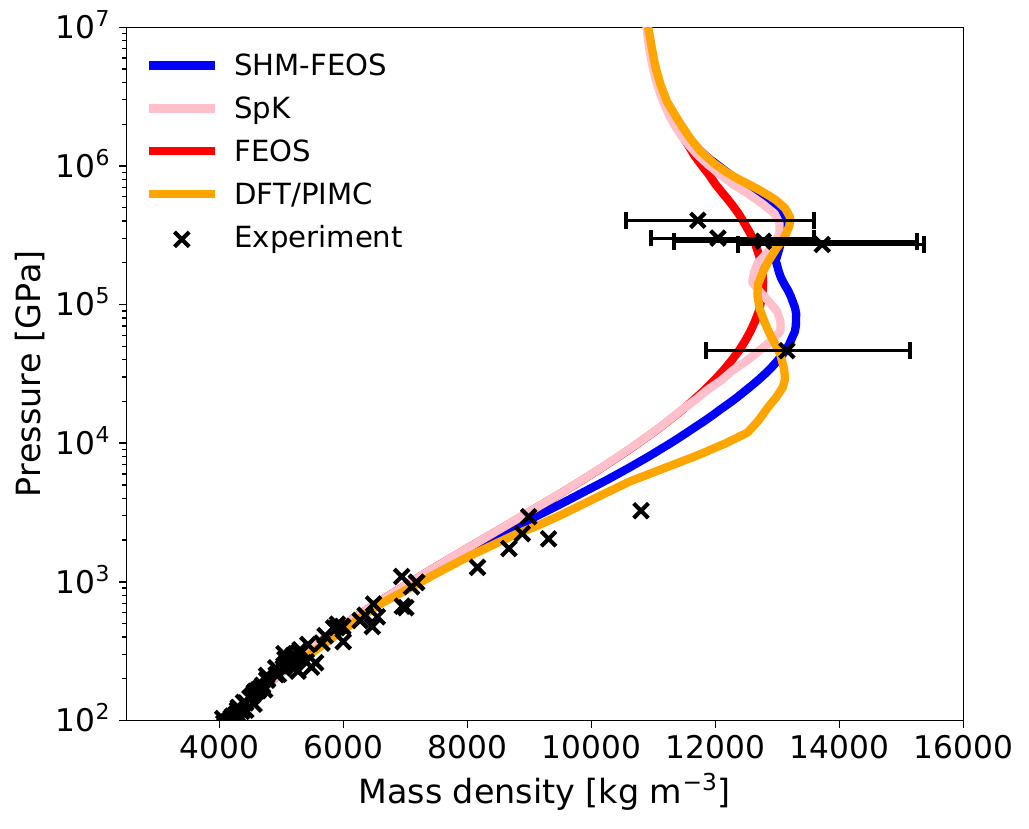}
    \caption{
        Principal Hugoniot data for aluminium, comparing \SpK~(pink) against the \texttt{SHM-FEOS} model (blue), \texttt{FEOS} (red), the DFT/PIMC data of Militzer et al.~\cite{MilitzerPRE2021} (orange), and experimental data (crosses) from the Shock Wave Database \cite{TEOS}, with error bars from Vladimirov et al.~\cite{Vladimirov1984}.
    }
    \label{fig:Al_Hugoniot}
\end{figure}

We next look at aluminium, a material with an extensive set of experimental measurements that is expected to give two maxima in compressibility, due to the ionisation of both the K- and L-shells. In Fig.~\ref{fig:Al_Hugoniot}, principal Hugoniot data generated using \SpK, the \texttt{SHM-FEOS} model, \texttt{FEOS}, and from the ab initio data of Militzer et al.~\cite{MilitzerPRE2021} are compared against various experimental measurements on the Shock Wave Database \cite{TEOS}. Again, the lack of shell structure in the TF model means \texttt{FEOS} fails to capture the peaks in compressibility observed in the data by the other models. The agreement of both \SpK~and the \texttt{SHM-FEOS} data with the ab initio data is good for the K-shell, but both predict L-shell ionisation at much higher shock pressures. Further work is needed to understand the difference, but we believe it is reasonable to assume that models based on the SHM-$\ell$ will become less accurate the further an ion is from being hydrogenic. The experimental error bars for the highest pressure experiments are too large to draw meaningful conclusions about model accuracy from them. 

\section{Off-Hugoniot calculations} 
\label{sec:off_hugoniot}

Materials in real experiments traverse a wide range of physical conditions that lie off the Hugoniot. Accurate measurements of states reached in HEDP experiments are challenging to obtain, relying on the indirect inference of thermodynamic states through synthetic diagnostics and other observable quantities \cite{HatfieldNature2021, CrillyPoP2018, CrillyPoP2020, ChittendenPoP2016}. The active field of first-principles molecular dynamics simulations provides state-of-the-art methods for obtaining information about the states of matter, particularly in warm dense matter (WDM) conditions where many of the approximations used in simpler theories break down \cite{MurilloPRE2010, GaffneyHEDP2018}. A plethora of experimental and computational studies exist exploring exotic states of matter in theoretically challenging conditions \cite{BonitzPoP2020, DornheimPoP2017, DornheimPR2018, DornheimPoP2023, FalkHPL2018, FalkPRL2018, RileyIoP2021}. 

\begin{figure}[!ht]
    \centering
    \includegraphics[width=0.49\textwidth]{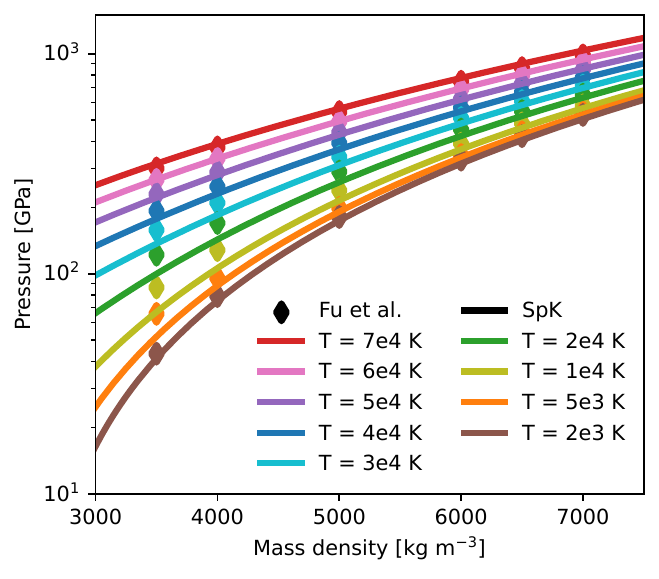}
    \caption{
        Aluminium pressure along a number of isotherms, comparing \SpK~(solid) against the ab initio molecular dynamics data of Fu et al.~(diamonds) \cite{FuPoP2017}.
    }
    \label{fig:Al_ab_initio}
\end{figure}

We limit the scope of off-Hugoniot examples to exploring a range of conditions for two example materials: aluminium, due to its frequent use in wire array implosions at Imperial College \cite{JenningsPoP2010, HallidayPoP2022}; and a 50:50 mix of deuterium and tritium (DT), which experiences quasi-isentropic compression following the deliverance of the final shock pulse to targets in ICF experiments. In Fig.~\ref{fig:Al_ab_initio}, we compare the pressure of aluminium calculated using \SpK~along isotherms against the ab initio data of Fu et al.~\cite{FuPoP2017}. We observe good agreement between both models, particularly at higher densities and across the entire isotherms at the lowest and top few temperatures. The lowest density points at $\sim10^{4}$ K temperatures lie firmly within the WDM regime and see the largest disagreement, at most with \SpK~predicting pressures $23\%$ below the ab initio data.

\begin{figure}[!ht]
    \centering
    \includegraphics[width=0.48\textwidth]{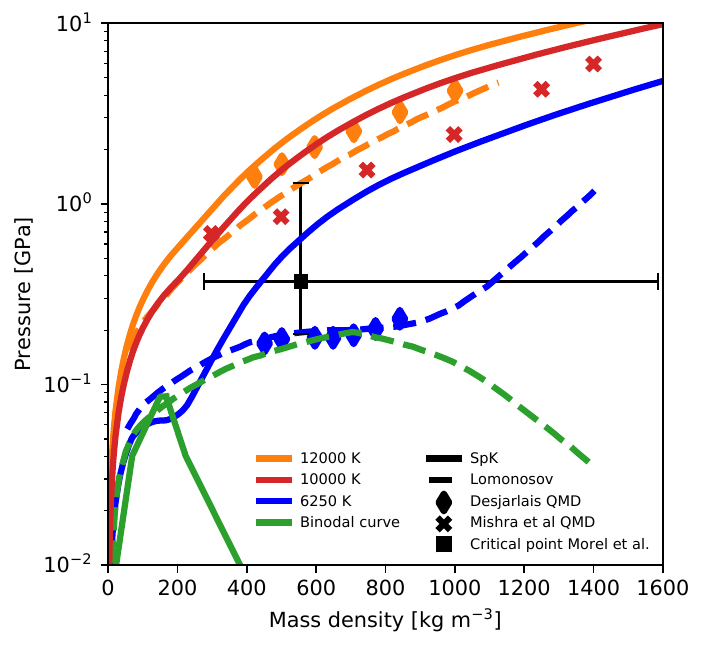}
    \caption{
        Aluminium pressure produced by \SpK~(solid) along low-density isotherms, compared to the model EoS of Lomonosov (dashed) and QMD simulation data from Desjarlias \cite{LomonosovLPB2007} (diamonds) and Mishra et al.~\cite{MishraJoP2012} (crosses). The critical point (with error bars) is taken from the review of Morel et al.~\cite{MorelIJT2009} and is identical to the point shown in Fig.~\ref{fig:Al_Maxwell}. For \SpK~and Lomonosov's model, the binodal curves bounding the liquid-vapour coexistence region are also shown in green. 
    }
    \label{fig:Al_pressure_low_density_isotherms}
\end{figure}

In wire array implosions, the magnetic pressure acting radially inwards is balanced by the thermal pressure. Due to the plasma temperatures achieved, the densities on-axis are below solid density. Though, the states achieved are non-thermal \cite{HallidayPoP2022, Niasse2012Thesis}, an understanding of the accuracy of the EoS in WDM states in thermal equilibrium is still of value. In Fig.~\ref{fig:Al_pressure_low_density_isotherms}, three pressure isotherms compare \SpK~with the model of Lomonosov \cite{LomonosovLPB2007}, and the QMD data of Desjarlais \cite{LomonosovLPB2007} and Mishra et al.~\cite{MishraJoP2012}. The binodal curve of Lomonosov's model is also compared with that of \SpK. The critical point data from Morel et al.~\cite{MorelIJT2009}, is again shown. When generating the data, a brief effort was made to tune free model parameters to reach good agreement with the critical temperature from Morel et al.~\cite{MorelIJT2009}. However, the critical pressure and density of \SpK~lie outside the error bars of Morel et al. Compared to Fig.~\ref{fig:Al_ab_initio}, greater disagreement is observed along the isotherms, most notably at 6250 K and $\sim800~\text{kg m}^{-3}$ where differences reach almost an order of magnitude. Moreover, a substantial disagreement is seen between the binodal curves above $\sim200~\text{kg m}^{-3}$. This is the lower density boundary of the interpolation region between the modified Saha and TF models.

The differences in the extent to which \SpK~agrees with ab initio data in compressed or rarefied states highlight two obvious areas in need of improvement. Firstly, a more sophisticated method of interpolating between the modified Saha and TF models is required. Secondly, the semi-empirical bonding corrections and present IPD models are too crude to capture the complex interaction physics that occur in WDM conditions. 

\begin{figure}[!ht]
    \centering
    \includegraphics[width=0.48\textwidth]{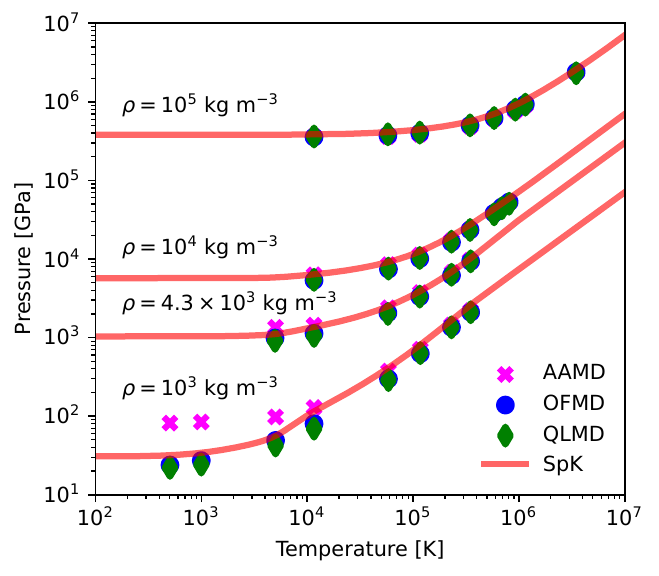}
    \caption{
        DT pressure along a number of isochores relevant to ICF. \SpK~(red curves) is compared against AAMD (magenta crosses), OFMD (blue circles), and QLMD (green diamonds) simulations of Kang et al.~\cite{KangMRE2020}.
    }
    \label{fig:DT_ab_initio}
\end{figure}

A material fundamental to fusion research is the 50:50 mixture of deuterium and tritium (DT). There is little experimental validation data available for DT in the high-pressure regime, so we benchmark \SpK~for this material by comparing against the molecular dynamics (MD) data of Kang et al.~\cite{KangMRE2020}. In Fig.~\ref{fig:DT_ab_initio}, we compare the pressure predicted by \SpK~against their average atom (AAMD), orbital-free (OFMD), and quantum Langevin (QLMD) MD simulations for a range of conditions relevant to ICF. \SpK~data closely matches the majority of the MD simulations, agreeing more with the OFMD and QLMD data in the lowest temperature and density conditions where the AAMD data differs most. Kang et al.~attribute this disagreement to the lack of correlation effects in the AAMD calculations \cite{KangMRE2020}.

In the above examples, \SpK~performs well compared to more sophisticated methods for compressed states and less so in WDM conditions. Given the very low computational cost of our approach, such results provide encouraging examples of where in parameter space global EoS tables generated by \SpK~are most accurate, alongside clear avenues for improvement.

\section{1D indirect-drive capsule simulations: model sensitivity comparison}
\label{section:ICF}

The motivation for improving global microphysics models is to enable higher-fidelity modelling of HEDP experiments. In this section we apply EoS and opacity tables generated using \SpK~in the integrated modelling of ICF capsule implosions and assess the sensitivity of the results to the choice of EoS model. We utilise the Eulerian RMHD code, \texttt{Chimera}, with $P_{1/3}$ \cite{OlsonJQSRT2000} multigroup \cite{ChittendenPoP2016} automatic flux-limited radiation transport \cite{StoneAJSS1992}, flux-limited Spitzer-H\"arm thermal transport \cite{SpitzerPR1953} with explicit super-time-stepping \cite{MeyerJCP2014}. The code also has extensive extended MHD capabilities \cite{WalshPoP2020} and Monte-Carlo alpha particle transport \cite{TongNF2019}, though these are not utilised in the present study. In the simulations presented here, the thermal conductivity is calculated from using the average charge state predicted by \SpK. Therefore, when \SpK~EoS data is utilised, all closure relations consumed by the simulations are obtained using a consistent microphysical description.

\begin{figure}[!ht]
    \centering
    \includegraphics[width=0.49\textwidth]{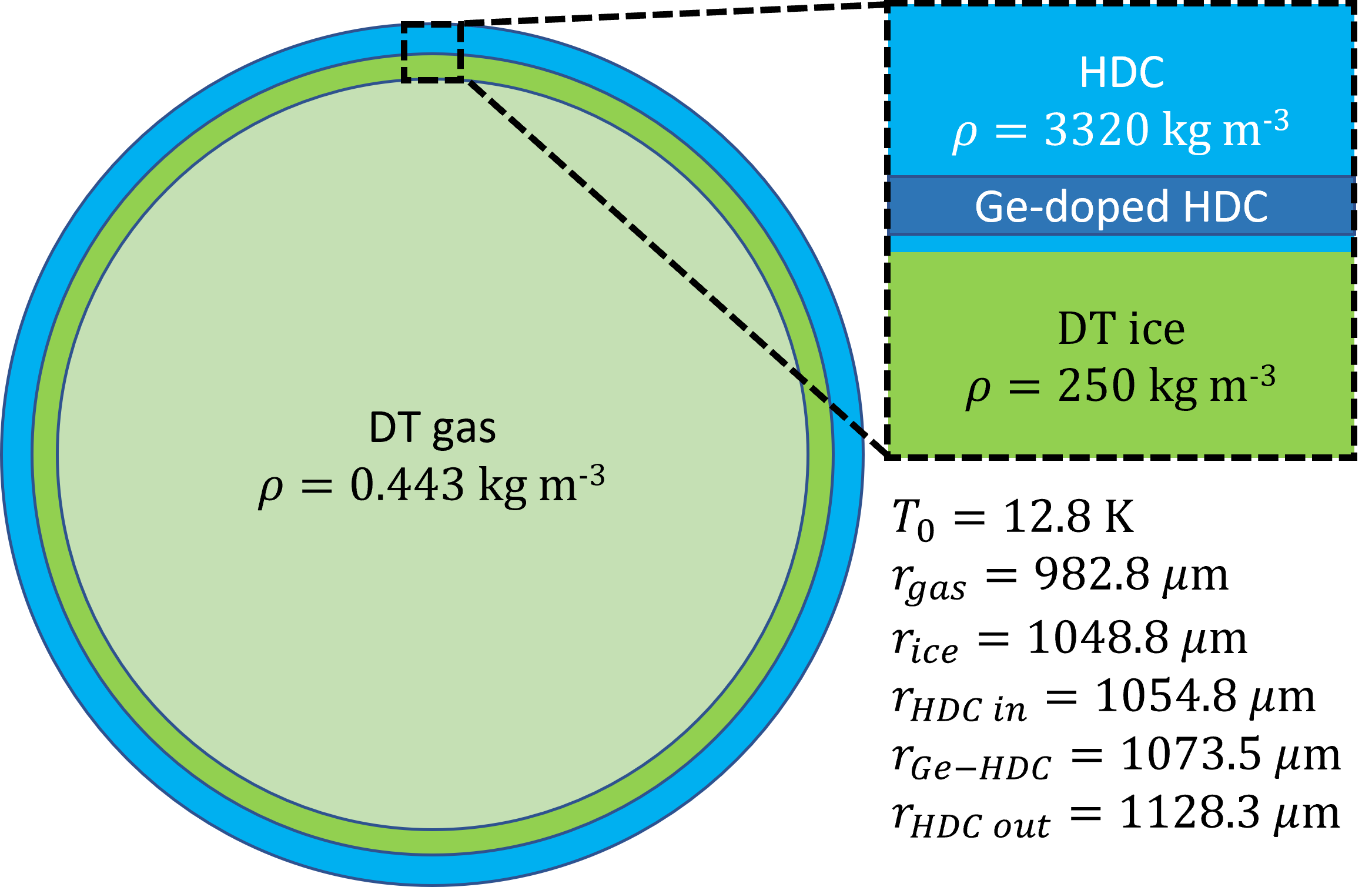}
    \caption{
        The schematic of the capsule and initial conditions of the materials used as input to the simulations in this work.
    }
    \label{fig:210808_capsule}
\end{figure}

The present study is on 1D capsule-only simulations of an indirect drive implosion, with the capsule and radiation drive based on the N210808 shot at the National Ignition Facility (NIF) \cite{AbuShawarebPRL2022, KritcherPRE2022, ZylstraNature2022}. The target schematic and initial conditions are given in Fig.~\ref{fig:210808_capsule}, with the tungsten layer replaced with germanium scaled to the atomic mass of tungsten. A time-varying, frequency-dependent radiation drive profile obtained from a hohlraum calculation \cite{Weber2021} was used as input for the radiation energy density. Material opacities were all generated using \SpK. When \texttt{FEOS} provided all EoS data for the materials, the total power of the drive and the atomic fraction of the dopant layer were tuned by Crilly until agreement was seen between stagnation conditions from \texttt{Hydra} and \texttt{Chimera} simulations with alpha heating disabled. 

In Tab.~\ref{tab:N210808_sims}, we compare the impact on the simulations of changing the EoS models used to generate the data for the materials being simulated by analysing implosion performance metrics. For the high-density carbon (HDC), \SpK, \texttt{FEOS}, and the \texttt{SHM-FEOS} model were used to generate the tabulated data. For the DT, only \SpK~and \texttt{FEOS} were used, since the \texttt{SHM-FEOS} model can only currently be used to generate single-species materials. The doped layer is not accounted for in the HDC EoS, but is accounted for the in the opacity.

\begin{table*}[ht]
    \centering
    \renewcommand{\arraystretch}{1.5}
    \begin{tabular}{c c c c c c c} 
    \hline\hline
    DT EoS & HDC EoS & $v_{\text{max}}$ [km/s] & $t_{\text{BT}}$ [ns] & $(\rho R)_{\text{BT}}$ [kg/m$^{2}$] & $\langle T_{i}\rangle_{\text{BA}}$ [keV] & $Y_{\text{total}}$ [kJ]\\ [0.5ex]  
    \hline
    \texttt{FEOS} & \texttt{SHM-FEOS} & 412.6 & 9.11 & 9.83 & 4.23 & 49.5\\ 
    \texttt{FEOS} & \texttt{FEOS} & 421.7 & 8.99 & 9.06 & 4.34 & 52.3\\
    \texttt{FEOS} & \SpK~& 416.3 & 9.09 & 9.59 & 4.29 & 50.3\\
    \SpK~& \texttt{FEOS} & 423.8 & 8.95 & 9.06 & 4.36 & 54.8\\
    \SpK~& \SpK~& 418.9 & 9.06 & 10.64 & 4.32 & 53.1 \\ 
    \hline\hline
    \end{tabular}
    \captionsetup{width=0.9\textwidth, justification=centering}
    \caption{
        Parameters of interest for comparing the simulation sensitivity to EoS model choice. From left to right is the DT EoS, HDC EoS, maximum fuel-mass-averaged implosion velocity, bang time, hot spot areal density at bang time, burn-averaged ion temperature, and total energy yield (burn off).
    }
    \label{tab:N210808_sims}
\end{table*}

\begin{table*}[!ht]
    \centering
    \begin{tabular}{ c c | c c c | c c c } 
     \hline\hline
     &&\multicolumn{3}{c|}{Shock in HDC} & \multicolumn{3}{c}{Shock in DT ice}\\
     \hline
     DT EoS & HDC EoS & $T$ & $p$ & $u_{s}$ & $T$ & $p$ & $u_{s}$\\
     \hline
     \texttt{FEOS} & \texttt{SHM-FEOS} & 2.55 & 1330 & 28.1 & 2.72 & 202 & 34.4\\ 
     \texttt{FEOS} & \texttt{FEOS} & 2.81 & 1420 & 29.1 & 2.91 & 230 & 35.7\\
     \texttt{FEOS} & \SpK~& 2.51 & 1320 & 28.0 & 2.66 & 199 & 34.2\\
     \SpK~& \texttt{FEOS} & 2.81 & 1420 & 29.1 & 2.01 & 205 & 32.9\\
     \SpK~& \SpK~& 2.51 & 1320 & 28.0 & 1.79 & 183 & 31.3\\ 
     \hline\hline
    \end{tabular}
     \captionsetup{width=0.9\textwidth, justification=centering}
    \caption{
        The states behind the first shock propagating through the HDC and DT ice layer in the simulations.
    }
    \label{tab:shock_data}
\end{table*}

Immediate correlations can be seen between higher implosion velocities, earlier bang times, and the burn-averaged ion temperature. The former correlation is explained by higher velocity implosions converging to stagnation pressures faster, and the latter because there is more kinetic energy to convert into hot spot internal energy. The SHM-$\ell$-based HDC tables have peak implosion velocities below those of the TF-based \texttt{FEOS} data. On the other hand, using \SpK~to generate the DT EoS appears to increase the peak velocity, though with less sensitivity compared to the HDC EoS. To understand this sensitivity and the coupling to the other performance metrics, we examined the shock timings and obtained the states behind the shock front by examining distance-time plots. 

The downstream pressure, temperature, and the shock velocity behind the first shock driven into the ablator and then the same shock entering ice layer are given in Tab.~\ref{tab:shock_data}. The simulations with EoS data for HDC generated using \texttt{FEOS} have higher shock pressures, temperatures, and shock velocities in the HDC. However, the downstream HDC state will lie firmly in the TF region of the \SpK~table, so such a sensitivity would not be expected following a shock of this strength.

\begin{figure}[!ht]
    \centering
    \includegraphics[width=0.49\textwidth]{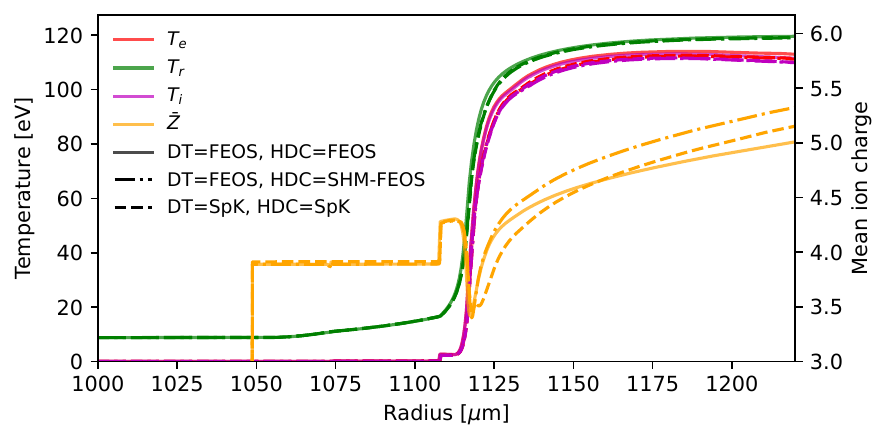}
    \caption{
        Radial profiles of the average ionisation (orange) and the temperatures of the electrons (red), ions (magenta) and radiation (green) of a section of the capsule during the ablation phase. Simulations where \texttt{FEOS} generates both EoS tables (solid) are compared against those where \SpK~generates both (dashed) and where the \texttt{SHM-FEOS} model generates the HDC data (dot-dashed).
    }
    \label{fig:210808_ablation}
\end{figure}

To understand this, we examine the radial profiles during the ablation phase, with Fig.~\ref{fig:210808_ablation} showing the mass density and average ionisation. The simulations that mix \texttt{FEOS} and \SpK~are ignored as the DT EoS does not impact this stage of the implosion. The ablated carbon is more ionised when SHM-$\ell$-based EoS models are used to generate the HDC data. This suggests more of the radiation energy is being absorbed in the ablated plasma and acting to ionise the K-shell in these simulations, instead of raising the thermal energy, leading to less energy being transported to the dense HDC and a weaker shock being driven. This is consistent with the raised specific heat capacity expected when beginning to ionise the K-shell, predicted to occur in the ablated plasma temperatures. This result suggests that it is not just the accuracy of the Hugoniot that is important for predicting shock timings, but also the coupled microphysics of the ablated plasma that must be accurate too.

\begin{figure}[!ht]
    \centering
    \includegraphics[width=0.49\textwidth]{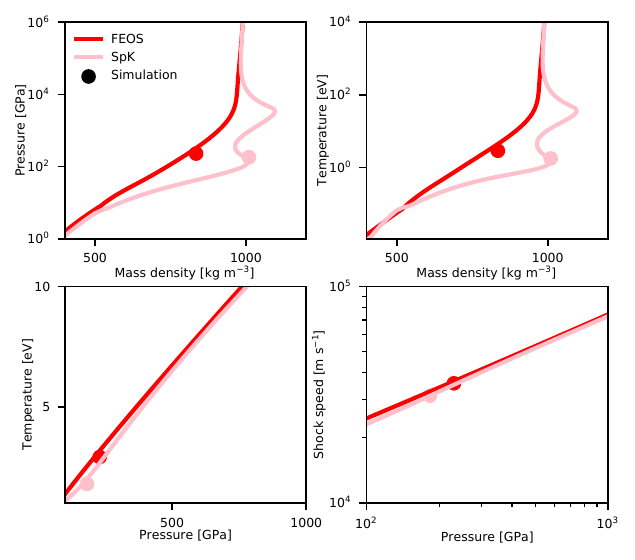}
    \caption{
        Hugoniot data for DT, comparing \texttt{FEOS} (red) and \SpK~(pink) data against the simulated downstream conditions in the DT ice (circles) from Tab.~\ref{tab:shock_data}.
    }
    \label{fig:DT_sim_Hugoniot}
\end{figure}

Another correlation in Tab.~\ref{tab:shock_data} is the temperature and shock velocity in the DT ice are consistently lower when the EoS is generated by \SpK, even with a higher drive pressure when \texttt{FEOS} generates the HDC EoS. This is understood by examining the DT Hugoniot predicted by both \SpK~and \texttt{FEOS} in Fig.~\ref{fig:DT_sim_Hugoniot}, including additional comparisons against both the temperature and shock speed. Included in the plots are the conditions behind the shock front for both simulations, revealing on-Hugoniot states and pressures coinciding with the increase in compressibility and specific heat capacity due to molecular dissociation in the \SpK~data. Therefore, less of the shock energy from the drive is partitioned into raising the thermal energy. The increase in compressibility also corresponds to a lower shock velocity. This reduced DT ice temperature could explain the increase in peak implosion velocity, since the adiabat would be lower, thus increasing the efficiency of conversion of drive energy to overall kinetic energy after the later shocks are driven. The opposite relationship being observed in the HDC can be explained as a result of less energy being coupled to the target when \SpK~generates the HDC EoS, which dominates over the reduced adiabat.

\begin{figure}[!h]
    \centering
    \includegraphics[width=0.49\textwidth]{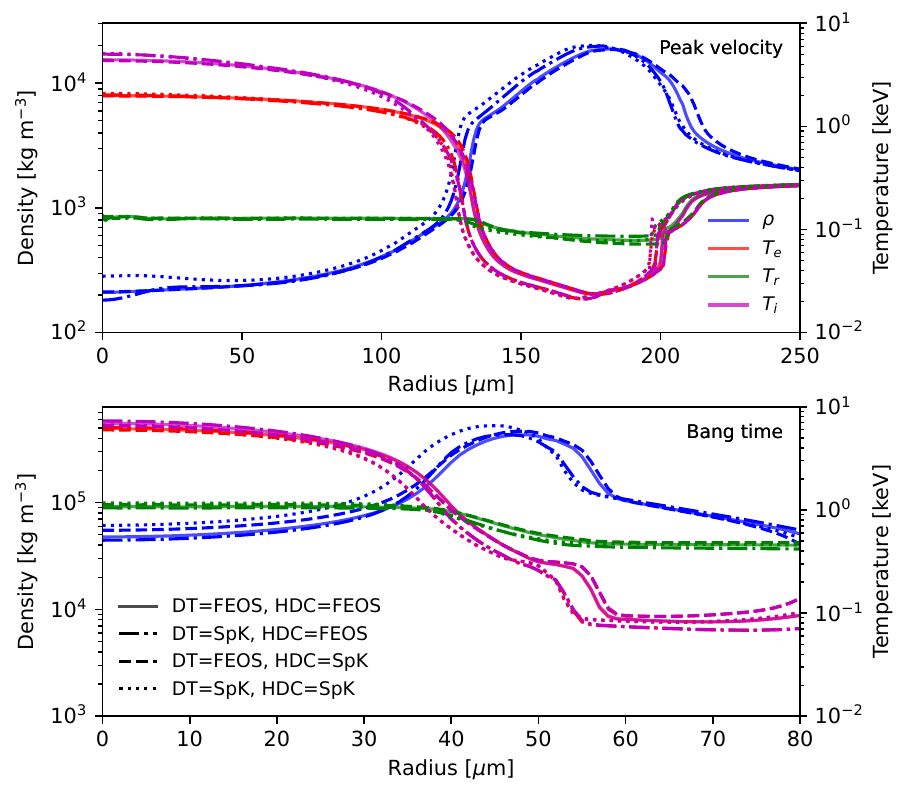}
    \caption{
        Radial profiles of the mass density (blue) and the temperatures of the electrons (red), ions (magenta) and radiation (green) at times of peak velocity (top) or bang time (bottom). The line styles indicate which model generated the EoS for each material.
    }
    \label{fig:210808_profiles}
\end{figure}

To understand the more integrated performance metrics in Tab.~\ref{tab:N210808_sims}, the radial profiles of the mass density, the electron, ion, and radiation temperatures are given at times of peak velocity and at bang time in Fig.~\ref{fig:210808_profiles}. Each of these events occurs at a slightly different time in the different simulations. We ignore where the HDC EoS is generated using the \texttt{SHM-FEOS} model is ignored here, since the behaviour is qualitatively similar to when \SpK~generates the data. The highest compression can be seen when \SpK~generates the EoS for both materials, likely due to both the added compression by the effect of molecular dissociation in the ice layer, but also due to the lower adiabat of the drive. The simulation where the DT and HDC EoS data were made by \SpK~and \texttt{FEOS}, respectively, shows similarly high compression at peak velocity, but then a low-density, high-temperature hot spot and a thinner ice layer at bang time. The maximum velocity is highest for this configuration in this comparison, though, as can be seen in Tab.~\ref{tab:N210808_sims}. High implosion velocities result in greater heating and the final compression being less isentropic, stagnating the implosion earlier and at larger radii \cite{Atzeni2009}, as can also be seen in Fig.~\ref{fig:210808_profiles}. This effect is likely dominant over the added compressibility of the \SpK-generated DT EoS, leading to the comparatively low areal densities in simulations where \texttt{FEOS} built the HDC EoS. 

The overall yield will be raised by higher areal densities and as a stronger function of temperature \cite{SpearsPoP2012}. The two implosions with \texttt{FEOS} providing the HDC EoS have the highest temperatures, but the simulation with \SpK~providing the EoS for both materials has the second highest yield, due to the significantly higher $(\rho R)_{\text{BT}}$. 

\section{Discussion and Future Work}
\label{section:discussion}
	
In this work we have described the new EoS capabilities of \SpK, extending the existing atomic physics, non-LTE, and opacity capabilities reported by Crilly et al.~\cite{CrillyHEDP2023}. We have combined the robust models utilised in \texttt{FEOS} with the fast statistical methods underpinning the atomic model in \SpK, creating a global model that captures atomic shell structure in weakly coupled conditions (where the EoS and opacity data are self-consistent). The Maxwell construction enables the modelling of materials from their empirical, zero-pressure reference conditions through the liquid-vapour coexistence region into gaseous and, ultimately, plasma states. By implementing a model for diatomic molecular dissociation, materials fundamental to fusion experiments, such as $D_{2}$ and DT can be modelled. 

Hugoniot calculations in Section \ref{section:Hugoniot} demonstrate the significant improvements over \texttt{FEOS} that are offered by \SpK's inclusion of electron shell structure and dissociation, if relevant. We show good agreement with well-known models, experimental data, and reasonable agreement with more sophisticated, expensive methods. \SpK~performs very well in benchmarks against QMD data for compressed, off-Hugoniot states in Section \ref{sec:off_hugoniot}, but more substantial disagreement is observed in WDM conditions below solid density. The analysis in Section \ref{section:ICF} demonstrates that \SpK~can be effectively utilised in integrated simulations and highlights sensitivities that can be explained by the physical differences between the models. Given the tuning of the drive and dopant fraction to match \texttt{Hydra} simulations using \texttt{FEOS} EoS data only, no meaningful conclusions can be drawn about which results are more correct. However, by highlighting the sensitivity to microphysics model choices, we evidence the need for more accurate global models to reduce errors in predictive capabilities.

There are numerous further developments that could improve \SpK's predictive capabilities. We intend to replace the cubic spline method with a more sophisticated, thermodynamically consistent interpolation between the TF and modified Saha regions of the table. By doing so, the EoS data would be thermodynamically consistent everywhere in the table. The present contributions to the EoS made by IPD are limited to the DH and IS models. By extending the existing capability to calculate IPD through the static structure factor approach \cite{CrillyHEDP2023} into a self-consistent contribution to the EoS, we can improve the accuracy of the EoS at moderate coupling. The \lq{Debye charging\rq}~process \cite{DornheimPR2018, IchimaruPR1987, LevinBJP2004} is a promising option for converting IPD into an equivalent contribution to the Helmholtz free energy. Extending these methods to include calculations of the dynamic structure factor could be utilised obtain self-consistent estimates of the plasma microfield distribution, which could replace the Q-fit the model for reducing the statistical weights of bound states. The EoS could also be extended to enable non-LTE calculations from the existing capabilities \cite{CrillyHEDP2023}. Correlation effects could be incorporated through the parametrisation of the corresponding Helmholtz free energy by Groth et al. \cite{GrothPRL2017}, which will improve the estimates of the modified Saha region as the boundary of validity is approached.

There are a number of free parameters within the model that can be tuned to match validation data. These include the indices in the soft-sphere model \cite{FaikCPC2018}, the definition of loci that bound the interpolation region, and the definition of the effective temperature in the dissociation model (if applicable). Gaussian processes could be utilised to optimise the agreement of this data with validation data simultaneously, such as the critical point of aluminium in Fig.~\ref{fig:Al_Maxwell} and principal Hugoniot data.

By exploring calculations of the latent heat of melting across a range of densities, a model for high-pressure phase transitions could be implemented. Such capabilities would improve agreement with validation data, such as the principal Hugoniot data surrounding the melting of diamond in Fig.~\ref{fig:C_Hugoniot}. This could significantly improve the modelling of experiments sensitive to material strength, such as the launching of flyer plates driven by pulsed power \cite{FitzgeraldIEEE2023, FitzgeraldIJIE2024}. The benefit of electron shell structure could be extended into degenerate conditions by replacing the TF model with quantum average-atom approaches, such as those utilised in \texttt{Purgatorio} \cite{SterneHEDP2007}.

There are ongoing efforts at First Light Fusion (FLF) to develop the Transport and Microphysics code, \texttt{TRaMP}, which smoothly transitions between transport models in the regions they apply. By supplying ionisation equilibrium data from the atomic model, all microphysical properties could be calculated from the output of \SpK. 

Complimenting the RMHD codes developed at FLF are reduced models used to rapidly explore wide-ranging, multidimensional parameter spaces to obtain optimal configurations to explore in higher fidelity. \SpK~is used to provide the opacity data for the First Light Advanced Ignition Model (\texttt{FLAIM}) \cite{BellenbaumIFSA2024, SaufiIFSA2024} (this issue), a reduced model for volume ignition under ongoing development designed for such a purpose. Future studies using \texttt{FLAIM} will utilise the new EoS capabilities of \SpK~also.

\section*{Acknowledgements}

The authors would like to thank First Light Fusion for fully funding this work.
		
\bibliography{ARF_thesis}
\end{document}